# Probing Within Partially Coherent Microcavity Frequency Combs via Optical Pulse Shaping

**Fahmida Ferdous,**[1*] **Houxun Miao,**[2,3] **Pei-Hsun Wang,**[1] **Daniel E. Leaird,**[1] **Kartik Srinivasan,**[2] **Lei Chen,**[2] **Vladimir Aksyuk,**[2] **and Andrew M. Weiner**[1,4]

[1]*School of Electrical and Computer Engineering, Purdue University,465 Northwestern Avenue, West Lafayette, IN 47907-2035*
[2]*Center for Nanoscale Science and Technology, National Institute of Standards and Technology, 100 Bureau Dr, Gaithersburg, MD 20899, USA*
[3]*Nanocenter, University of Maryland, College Park, MD 20742, USA*
[4] *Birck Nanotechnology Center, Purdue University, 1205 West State Street, West Lafayette, Indiana 47907, USA*
[*]*fferdous@purdue.edu*

**Abstract:** Recent investigations of microcavity frequency combs based on cascaded four-wave mixing have revealed a link between the evolution of the optical spectrum and the observed temporal coherence. Here we study a silicon nitride microresonator for which the initial four-wave mixing sidebands are spaced by multiple free spectral ranges (FSRs) from the pump, then fill in to yield a comb with single FSR spacing, resulting in partial coherence. By using a pulse shaper to select and manipulate the phase of various subsets of spectral lines, we are able to probe the structure of the coherence within the partially coherent comb. Our data demonstrate strong variation in the degree of mutual coherence between different groups of lines and provide support for a simple model of partially coherent comb formation.

**1. Introduction:**

Optical frequency comb generation using a high quality factor (Q) microresonator has recently attracted significant interest due to its simplicity, and the small size, and compatibility with low-cost, batch fabrication processes of the microresonators. Frequency comb generation in a variety of micro systems including microtoroids [1-3], microspheres [4], microrings [5-10], and millimeter-scale crystalline resonators [10-15] has been demonstrated. Materials used in these systems include silica [1-4, 6], silicon nitride [5, 7-10], fused quartz [12], calcium fluoride [11, 15], and magnesium fluoride [10, 13, 14]. However, the physics behind the comb generation is not yet fully understood. The different types of comb formation are studied theoretically and experimentally [7, 10, 12, 13, 15]. Recently, time-domain studies based on optical pulse shaping [16] have revealed new information on the compressibility and the coherence of microresonator combs [7, 12]. Our previous work revealed two different paths to comb formation that lead to strikingly different time-domain behaviors [7]. Combs formed as a cascade of sidebands spaced by a single free spectral range (FSR) that spread from the pump (Type I) may be compressed to the bandwidth limit by using a pulse shaper to compensate the spectral phase. Such full compressibility provides evidence of stable spectral phase, i.e., coherence, analogous to that which is observed with mode-locked lasers. However, combs where the initial sidebands are spaced by multiple FSRs that then fill in to give combs with single FSR spacing (Type II) show degraded compressibility and partial coherence.

Figure 1 provides a schematic view of routes to comb generation. Figure 1(a) depicts the Type I comb generation process, in which initial comb lines are generated at ±1 FSR from the pump. Subsequent lines generated via cascaded four wave mixing spread out from the center. Energy conservation ensures that the new lines are exactly evenly spaced from the pump. Hence, this process is expected to result in a high degree of coherence.

Figure 1(b) shows the case where comb lines are initially generated at ± N FSRs from the pump, where N is an integer greater than one (N=3 for the example shown). Such behavior has been reported frequently in the literature. This process is again expected to produce equal frequency spacings, which should result in high coherence. Our data published in [7] demonstrate that both the process of Fig. 1(a) and the process of Fig. 1(b) admit high quality pulse compression, providing evidence of stable spectral phase and good coherence. By varying the pump laser, e.g., tuning it closer into resonance, additional spectral lines may be observed to fill in between those of Fig. 1(b), resulting in a Type II comb with nominally single FSR spacing. In the case illustrated in Fig. 1(c), each of the lines from Fig. 1(b) has spawned additional lines at ±1 FSR through an independent four-wave mixing process. Although the lines within any of the individual triplets resulting from this last process are expected to be equally spaced, there is nothing to guarantee that the spacings within the triplets are exact submultiples of the original multiple-FSR comb spacing: in fact, due to dispersion, it is very unlikely that these new frequency spacings will be exact submultiples of the spacing in Fig. 1(b). As a result of such an imperfect frequency-division process, the relative phases between certain groups of generated lines will vary with time. The time domain waveform, which arises from the interference of the generated optical frequency components, will therefore vary on a time scale set by the nonuniformity of the frequency spacings. In this sense we may say that the coherence is compromised.

One consequence of a process such as that illustrated in Fig. 1(c) would be that different groups of lines may exhibit different degrees of coherence. For example, the initial comb lines spaced by multiple FSRs, e.g., at frequencies {… -3, 0, 3 …} (relative to the pump in FSR units) in Fig. 1(b), are coupled and should therefore have equal frequency spacing and high mutual coherence. Likewise, individual triplets of lines, shown in Fig. 1(c), e.g., {-1, 0, 1}, which are spawned from a single line in the previous step, are coupled and may also exhibit relatively high mutual coherence.

In our experiments we test this hypothesis by using a pulse shaper to select such groups of lines out of a Type II comb. Time domain experiments confirm good compressibility and coherence. In additional experiments a pulse shaper is used to select other groups of lines for which direct coupling is not expected at the early stages of our model. In such cases we observe that compressibility and coherence are degraded.

With further evolution beyond that sketched, the various frequency triplets of Fig. 1(c) may each spawn additional lines through cascaded four-wave mixing, which will then approximately overlap. However, because of the imperfect frequency division, the overlap between newly generated lines will not be exact. RF beating measurements have recently been used to characterize the evolution of such Type II combs [10]. The beating data provide evidence that individual comb lines may exhibit spectral substructure too fine to be resolved in normal optical spectrum analyzer measurements, consistent with the picture suggested here. Such spectral substructure leads to increased noise if the now roughly, but not exactly, overlapping frequency components fail to lock together. Increased RF noise in the Type II regime was also reported in [12] ("L2" regime in their notation). In [12] further tuning of the pump into the resonance led to a new regime ("L3" in their notation) characterized by substantial narrowing of the RF spectrum, with improved noise performance. Reduction of the RF noise with further tuning into resonance was also reported in [8, 10, 17]. The experiments presented in the current paper focus on the Type II regime and result in the new observation, not seen in previous studies, that some subgroups of comb lines retain mutual coherence markedly higher than that of the overall partially coherent Type II comb. Such coherence behavior contains structure that bears on the early stages of the comb generation process

## 2. Experimental Procedure:

Our experiments use a silicon nitride ring resonator with 100 μm outer radius, 2 μm waveguide width and 550 nm waveguide thickness for the frequency comb generation. Light is coupled into/out of the resonator via an on-chip waveguide with 1 μm width and an 800 nm ring-waveguide gap. The two ends of the waveguide are inversely tapered to 100 nm for low loss waveguide-fiber coupling (1.5 dB per facet with lensed fiber). The intrinsic Q of the resonator is estimated to be $3.0 \times 10^6$. The average FSR for the series of high Q modes is measured to be ≈ 1.85 nm.

Strong CW pumping light (estimated to be 27.6 dBm into the accessing waveguide) is launched into an optical mode at around 1551.6 nm. The generated frequency comb is sent to a line-by-line pulse shaper as shown in Fig. 2, which both assists in spectral phase characterization and enables pulse compression [7]. By slowly tuning the wavelength of the pump light from the blue to the red, we first observed an optical frequency comb with spectral lines spaced by 3 FSRs; then the cavity modes in between fill in to form a comb with one FSR spacing. Figure 3(a) shows the spectrum of the 3 FSR spacing comb with the pump line attenuated by 15 dB by the pulse shaper. Figure 3(b) shows the autocorrelation traces before (blue) and after (red) phase compensation by the spatial light modulator in the pulse shaper. Prior to compensation the waveform is only weakly modulated in time, while after compensation a clean train of isolated pulses is obtained.  The autocorrelation trace calculated for the spectrum shown in Fig. 3(a) assuming flat spectral phase is also plotted (black). The measured autocorrelation trace after phase compensation is in good agreement with the calculation which is based on a stable, deterministic spectral phase function. This agreement is a signature of coherent behavior. Figures 3(c) and 3(d) show the spectrum and the autocorrelation traces of the corresponding 1 FSR comb (after slowly tuning the laser wavelength to the red side). The significant difference of the measured autocorrelation trace from the calculated trace (again computed using the measured optical spectrum analyzer (OSA) spectrum and assuming flat spectral phase) indicates partial (i.e., degraded) coherence for this Type II comb.  The increased level of the autocorrelation in between the main peaks is especially relevant, as on-off contrast is the key observable that differentiates between the intensity autocorrelation of a coherent ultrashort pulse and the intensity autocorrelation of broadband noise [18, 19].

**3. Experimental Results for Probing the Partial Coherence Properties of a Type II Comb:**

To probe more deeply into the partial coherence properties of a Type II comb, we use a pulse shaper to select out various subsets of lines from the spectrum shown in Fig. 3(c) via amplitude filtering.  The phase corrections applied to each comb line are kept the same as in the previous phase compensation experiment (which resulted in the red trace in Fig. 3(d)). Filtered spectra and corresponding autocorrelation traces are shown in Fig. 4. Figure 4(a) shows the spectrum resulting when every third comb line is selected, including the pump. This leaves lines {… -6, -3, 0, 3, 6 …}, which is the same set of comb lines present in the original multiple FSR spectrum of Fig. 3(a).  The autocorrelation data are similar to those of Fig. 3(b). Without phase compensation the autocorrelation is modulated only weakly; phase compensation results in strong compression. As before, an autocorrelation is also calculated for comparison by taking the measured spectrum and assuming flat spectral phase. The very close agreement between experimental and calculated autocorrelations, including on-off contrast, shows that high coherence is maintained for this subfamily of lines.

   Figures 4(c) and 4(d) show the results when a group consisting of every third line is selected, but this time shifted from the pump. In particular, the pulse shaper is programmed to pass lines { … -5, -2, 1, 4 …} (frequencies relative to the pump in FSR units). These lines are not directly coupled in the early stages of comb formation outlined in Figs. 1(b)-(c). Figures 4(e) and 4(f) show the results of another family with lines { … -4, -1, 2, 5 …}.  Now there is significant difference between experimental and calculated (again assuming flat phase) autocorrelations. For the data shown in Figs. 4(e-f), the spectrum is relatively smooth. Therefore the autocorrelation calculated for full coherence and flat phase has high on-off contrast.  In this case the loss of contrast in the experiment is seen very prominently. Phase compensation provides only weak compression, with the experimental autocorrelation minimum falling only to ≈ 20 % of the peak in Fig. 4(f). Clearly, the coherence is badly degraded. For the data of Figs. 4(c-d), the spectrum is uneven. As a result the calculated autocorrelation for the case of full coherence already has high background and low on-off contrast. Although additional loss of contrast is evident in the experiment, indicating reduced coherence, the appearance is less dramatic. The loss of mutual coherence for this set of lines will be supported with further data in Fig. 7 below.

Similar results are obtained in another experiment in which the pulse shaper is programmed to pass every second line. Data obtained for the set of lines including the pump {… -6, -4, 2, 0, 2, 4 …} are shown in Figs. 5(a) and 5(b); similar data are shown in Figs. 5(c) and 5(d) are obtained for the complementary set of alternating comb lines. Again poor autocorrelation contrast is observed, as expected since a majority of the selected lines are not directly coupled in the early stages of the presented model of comb formation.

We further probe the coherence within the comb in experiments in which sets comprising only three spectral lines are selected. For two spectral lines of fixed relative power, the intensity autocorrelation is independent of the relative phase. Therefore, three is the minimum number of lines which show sensitivity to spectral phase and its variations. Following a procedure similar to that reported in [12], the pulse shaper is used to vary the relative phase ΔΦ of the longest wavelength line, with the phases of the other two lines arbitrarily set to zero. Figure 6 shows the evolution of the autocorrelation traces with ΔΦ when the pulse shaper selects lines {-3, 0, 3}. These are a subset of the lines selected for the data of Figs. 4 (a) and (b) for which high coherence was observed. The traces vary substantially as the relative phase is changed, evolving from a single strong peak for ΔΦ=0 spaced by the autocorrelation period T to a pair of weak peaks spaced by T/2 for ΔΦ = π. Here the autocorrelation period T ≈ 1.43 ps is the inverse of the ≈ 700 GHz frequency separation between selected comb lines. When ΔΦ is increased to 2π, the autocorrelation recovers to the same form as ΔΦ=0. The experimental traces (colored lines) are in close agreement with those calculated (black lines) based on the measured spectrum and the programmed phases. The close correspondence between the traces shows that the phase fluctuations are small and the coherence among the selected lines is high. Quite different results are obtained for the triplet of lines {-2, 1, 4}, which is selected from the set of lines of Fig. 4(c) which showed degraded coherence. As shown in Fig. 7, the autocorrelation traces show only minor changes as ΔΦ is varied. This lack of dependence on the relative phase demonstrates that the coherence among this triplet of lines, which are not directly coupled in the early stages of our model of comb formation, is very weak. Similar results (not plotted) showing loss of coherence are observed when three lines are selected from the spectrum of Fig. 4(e).

Another way to view the results of such experiments is through the visibility curves V(ΔΦ), defined as [12]:

$$V(\Delta\Phi) = \frac{|p(\tau_p) - p(\tau_p + T/2)|}{|p(\tau_p) + p(\tau_p + T/2)|}$$

where $p(\tau_p)$ is the value of the intensity autocorrelation peak, $\tau_p$ is the delay at which the peak occurs, and $p(\tau_p+T/2)$ is the value of the intensity autocorrelation half way between the peaks. Note that because the autocorrelation is periodic for data sets such as those of Figs. 6 and 7, we have three equivalent measurement points representing the autocorrelation peaks [$p(\tau_p)$ and $p(\tau_p \pm T)$] and two equivalent measurement points [$p(\tau_p \pm T/2)$] representing the autocorrelation value midway between the peaks. In order to extract values for V from the data, for each autocorrelation trace we average the equivalent points representing $p(\tau_p)$ and $p(\tau_p+T/2)$, respectively, then plug into the equation given above. We estimate the statistical uncertainty by evaluating V using the various pairs of individual (not averaged) measurement points representing the autocorrelation at and midway between its peaks. Another means to assess the uncertainty is to compare the obtained values for *V(0)* and *V(2π)*, since the visibility should be unchanged with $2\pi$ phase shift. For six of the seven visibility curves reported below, the values for *V(0)* and *V(2π)* agree to within the estimated uncertainty. Experimental visibility curves are compared with ideal theoretical curves which are computed based on the measured optical power spectrum and assuming full coherence. Multiple traces of the optical spectra are recorded simultaneously with each autocorrelation measurement; this permits estimation of statistical uncertainty in the scale of the theoretical curves due to small variations in the measured spectra. Uncertainties quoted below represent one standard deviation.

Figure 8(a) shows visibility curves for triplets of lines separated by three FSRs. The visibility of the {-3, 0, 3} triplet (red line), which includes the pump, exhibits strong dependence on $\Delta\Phi$, with minimum visibility equal to 8 % ± 1 % of the maximum visibility. The data are close to the ideal theoretical curve (blue line). In contrast, the visibility curves for triplets shifted away from the pump, {-2, 1, 4} and {-7, -4, -1}, show only weak dependence on $\Delta\Phi$, with minimum visibility equal to 66 % ± 6 % and 77 % ± 7 % of the maximum visibility, respectively. For comparison, the ratio of minimum to maximum visibility that would be expected ideally (with full coherence) are 9 % ± 0.4 % and 24 % ± 1 %, respectively. (The difference in ideal visibility ratios is explained by differences in the relative intensities of the selected lines.) The strong reduction in visibility clearly signifies degraded coherence. Figure 8(b) shows visibility curves for triplets {-4, -2, 0} and {-3, -1, 1}, each of which is spaced nominally by 2 FSRs. Again only weak variation in visibility, hence low coherence, is observed. These data consistently show low coherence for line triplets which lack direct coupling in the early stages of our model of comb formation.

Figure 8(c) shows visibility curves for triplets spaced nominally by a single FSR. For the {-1 0 1} triplet, the visibility curve (red) exhibits large variation with $\Delta\Phi$, with minimum visibility equal to 13 % ± 13 % of the maximum visibility. This provides evidence that relatively good coherence remains, consistent with the coupling within this triplet of lines suggested by our model, Fig. 1(c). As a counter example, we next selected the triplet {-2 -1 0}, comprising the pump and two lines to one side of the pump. Since the -2 line would not have direct coupling to the -1 and 0 lines in the second step of our hypothesized comb generation model, degraded coherence is predicted. This is confirmed by the corresponding (black) visibility curve in Fig. 8(c), for which the variation is significantly reduced with minimum visibility equal to 58 % ± 5 % of the maximum visibility.

**4. Discussion:**

A final observation is that, although to a large extent, our data support the simple model suggested by Fig. 1(c) – groups of lines that are directly coupled should have high coherence, while groups of lines that are not directly coupled should lack coherence – the correspondence is not complete. For example, the peak visibility data for the {-1, 0, 1} triplet in Fig. 8(c) is somewhat lower than the theoretical trace, which suggests the coherence may be less than 100 %. Similarly, although variation of the visibility for the {-2 -1 0} triplet in Fig. 8(c) is clearly suppressed compared to the {-1, 0, 1} case, the suppression is incomplete: some variation remains. This suggests that although the coherence is reduced, it does not reach zero. An explanation for these observations may involve further cascaded four-wave mixing beyond the early stage of comb formation portrayed in Fig. 1(c), which will cause the individual subcombs to spread until they overlap, as discussed in [10] for both crystalline magnesium fluoride and planar silicon nitride microresonators. Figure 9 shows a sketch illustrating this process, in this case with the initial lines separated by 6 FSRs. The important point is that in the final stage illustrated, any region of the comb spectrum consists of distinct overlapped subcombs with slight differences in frequency and frequency spacing. Because the pulse shaper does not resolve the distinct subcombs, the overall coherence of selected comb lines depends on the simultaneous contributions of different subcombs. Simulations that model the generated frequencies as coherent within individual subcombs but incoherent across distinct subcombs are able to approximately reproduce our observations in the visibility data.

**5. Conclusion:**

In conclusion, we have used an optical pulse shaper to probe within a partially coherent frequency comb generated via cascaded four-wave mixing in a silicon nitride microresonator. Our measurements reveal a striking variation in the degree of mutual coherence exhibited for different groups of lines selected out of the full comb. The structure observed in the mutual coherence provides evidence consistent with a simple model of partially coherent comb formation.

This work was supported by the National Science Foundation under grant ECCS-1102110 and by the Air Force Office of Scientific Research under grant FA9550-12-1-0236. We gratefully acknowledge Jian Wang for the autocorrelator setup. Dr. Houxun Miao acknowledges support under the Cooperative Research Agreement between the University of Maryland and the National Institute of Standards and Technology Center for Nanoscale Science and Technology, Award 70NANB10H193, through the University of Maryland.

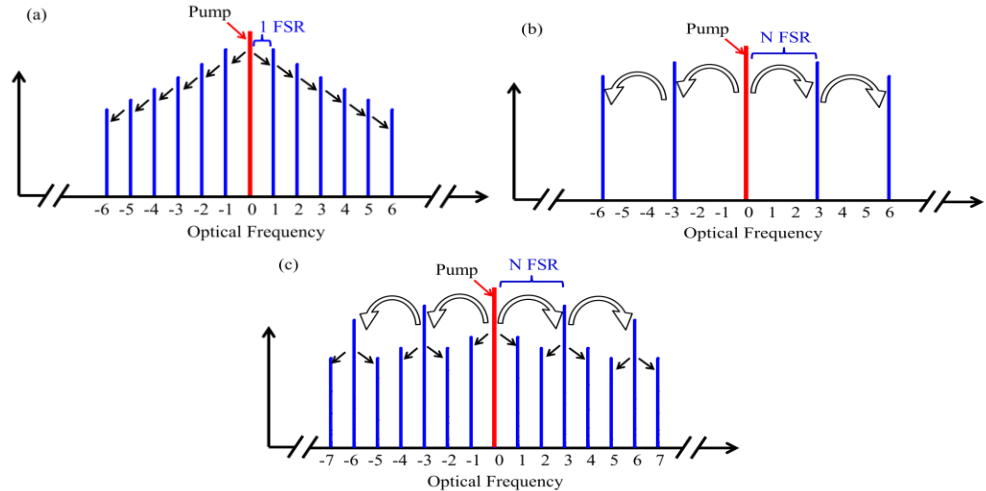

Fig.1. Possible routes to comb formation. The optical frequency axis is portrayed in free spectral range (FSR) units. (a) Case where initial comb lines are spaced by one FSR from the pump line, with subsequent comb lines, generated through cascaded four wave mixing, spreading out from the center. (b) Case where initial comb lines are spaced from the pump line by N FSRs, where N>1 is an integer. Here N=3 is assumed. (c) When the pump laser is tuned closer into resonance, additional lines are observed to fill in, resulting in spectral lines spaced by nominally 1 FSR.

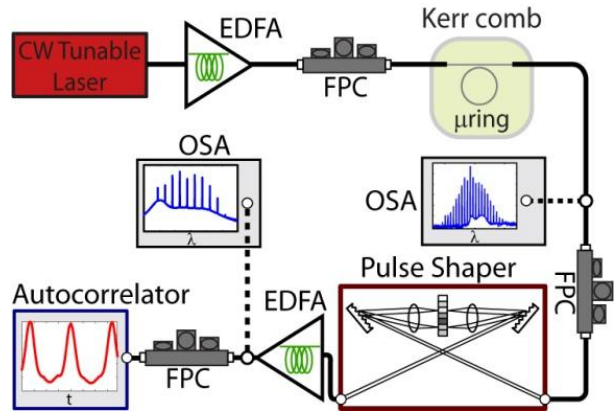

Fig.2. Schematic of the experimental setup for line-by-line pulse shaping of a frequency comb from a silicon nitride microring. CW: continuous-wave; EDFA: erbium doped fiber amplifier; FPC: fiber polarization controller; µring: silicon nitride microring; OSA: optical spectrum analyzer. The autocorrelator is constructed in a background-free, noncollinear geometry.

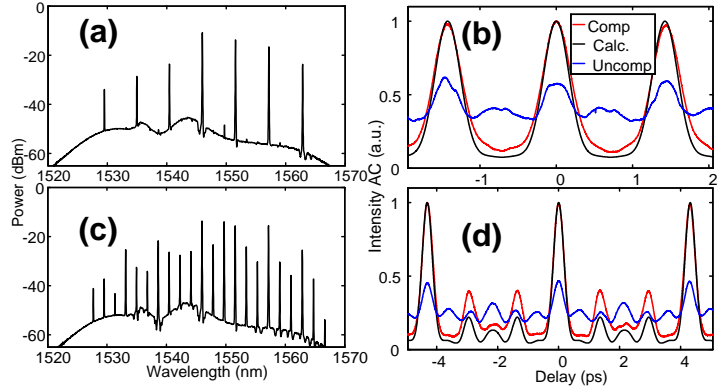

Fig.3. (a) Spectrum of the generated 3 FSR spacing comb after the pulse shaper. (b) Autocorrelation traces corresponding to (a). (c) Spectrum of the generated 1 FSR spacing comb after the pulse shaper. Here we tune the CW laser 53 pm further towards the red compared to (a). (d) Autocorrelation traces corresponding to (c). The pump line is attenuated by 15 dB by the pulse shaper. Blue and red traces are experimental traces before and after phase compensation respectively. Black traces are calculated by taking the OSA spectrums and assuming flat spectral phase.

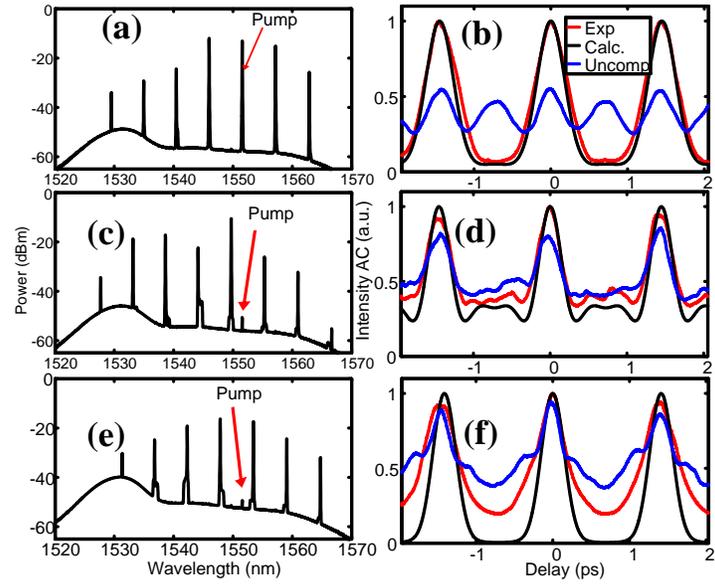

Fig.4. Spectra and autocorrelation traces for 3 subfamilies of comb lines with 3 FSR spacing selected from the spectrum shown in Fig. 3(c). Blue and red traces are experimental traces before and after phase compensation respectively. Black traces are calculated by taking the OSA spectrums and assuming flat spectral phase.

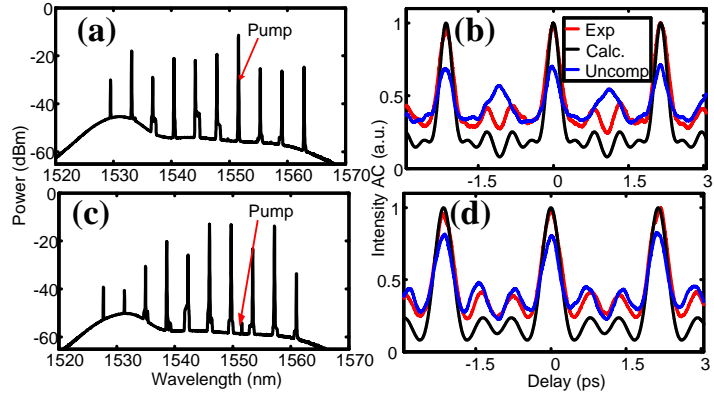

Fig.5. Spectra and autocorrelation traces for 2 subfamilies of comb lines with 2 FSR spacing selected from the spectrum shown in Fig. 3(c). Blue and red traces are experimental traces before and after phase compensation respectively. Black traces are calculated by taking the OSA spectrums and assuming flat spectral phase.

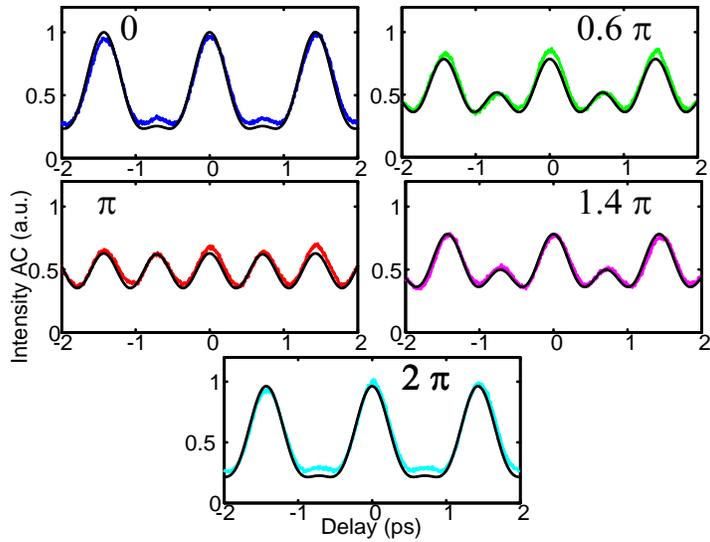

Fig.6. Autocorrelation traces for 3 line experiments for different $\Delta\Phi$ for lines {-3, 0, 3}, for which high coherence is observed. Here colored lines are the experimental traces and black lines are the simulated traces.

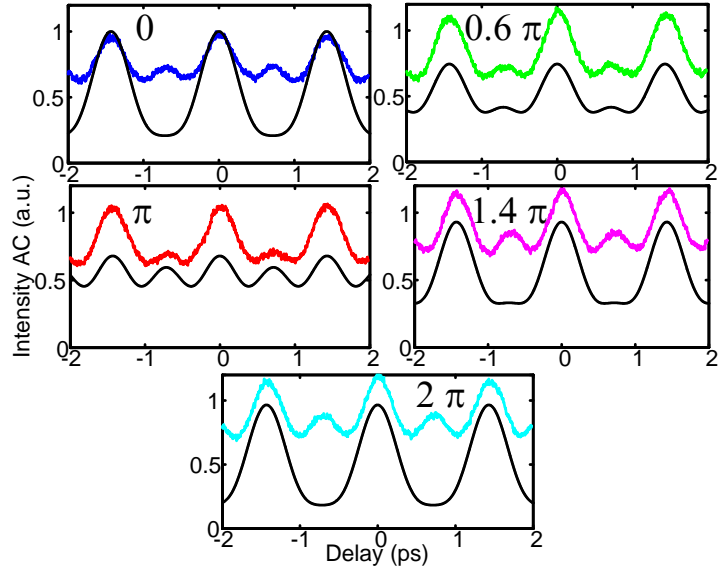

Fig.7. Autocorrelation traces for 3 line experiments for different ΔΦ for lines {-2, 1, 4}, for which low coherence is observed. Here colored lines are the experimental traces and black lines are the simulated traces.

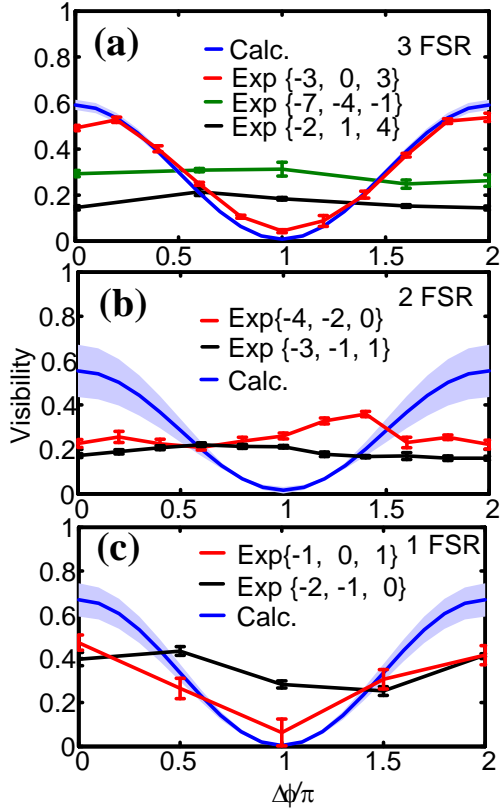

Fig.8. Visibility traces of (a) three subfamilies of 3 FSRs, (b) two subfamilies of 2 FSRs, and (c) two subfamilies of 1 FSR. Here in the visibility curves, red, green and black lines are the experimental data; blue lines are ideal theoretical curves calculated assuming full coherence based on the power spectra for the experimental cases for which high coherence is observed. Numbers in curly braces indicate the 3 lines that are used in the experiments. Error bars and shaded areas represent the mean ± one standard deviation.

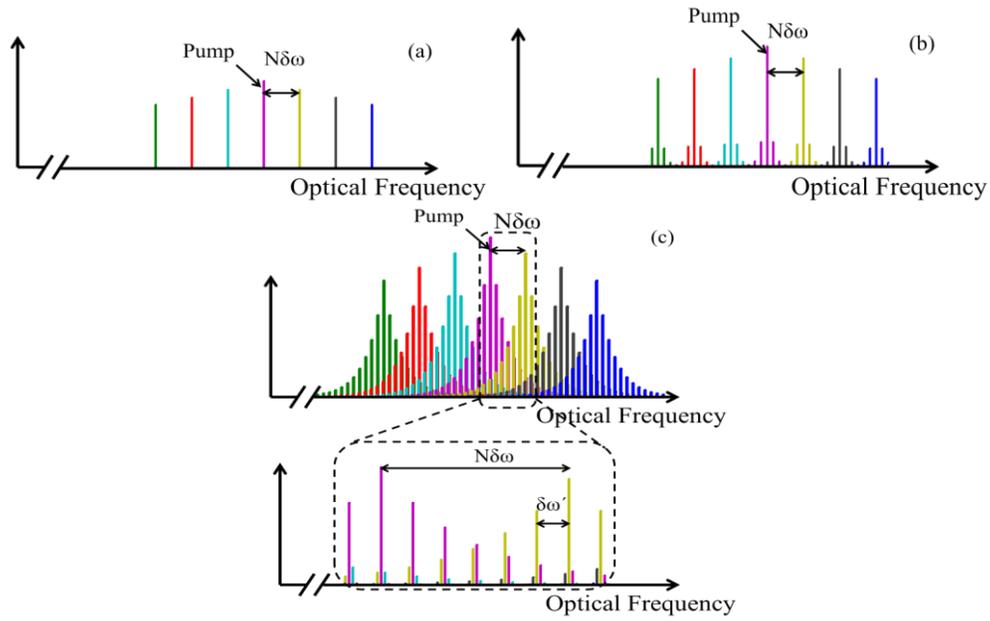

Fig.9. Proposed model for type II comb formation. (a) First: generation of a cascade of sidebands spaced by $N$ FSRs ($N\delta\omega$) from the pump. Here $N=6$ is illustrated. (b)-(c) The 2nd event is an independent four-wave mixing process, which creates new sidebands spaced by a different amount, $\pm n\delta\omega'$ ($n=1,2$ or 3....for different lines), from each of the lines in the previous step. Due to dispersion, it is very unlikely that the new frequency spacings will be exact submultiples of the original $N$ FSR spacing; i.e., $\delta\omega' \neq \delta\omega$.